\documentclass{iaus}

\usepackage{graphics,epsfig}

  \checkfont{eurm10}
  \iffontfound
    \IfFileExists{upmath.sty}
      {\typeout{^^JFound AMS Euler Roman fonts on the system,
                   using the 'upmath' package.^^J}%
       \usepackage{upmath}}
      {\typeout{^^JFound AMS Euler Roman fonts on the system, but you
                   dont seem to have the}%
       \typeout{'upmath' package installed. iaus.cls can take advantage
                 of these fonts,^^Jif you use 'upmath' package.^^J}%
      }
  \else
  \fi

  \checkfont{msam10}
  \iffontfound
    \IfFileExists{amssymb.sty}
      {\typeout{^^JFound AMS Symbol fonts on the system, using the
                'amssymb' package.^^J}%
       \usepackage{amssymb}%

      }{}
  \fi

  \IfFileExists{amsbsy.sty}
    {\typeout{^^JFound the 'amsbsy' package on the system, using it.^^J}%
     \usepackage{amsbsy}}
    {}

\newsavebox{\astrutbox}
\sbox{\astrutbox}{\rule[-5pt]{0pt}{20pt}}

\title[The Interplay among Black Holes, Stars and ISM in Galactic 
       Nuclei]{Blue colours of BL Lac host galaxies}

\author[J.K. Kotilainen \& R. Falomo]%
{J.K. Kotilainen$^1$ \and R. Falomo$^2$}

\affiliation{
$^1$ Tuorla Observatory, University of Turku, V\"ais\"al\"antie 20, 
21500 Piikki\"o, Finland \\[\affilskip]
$^2$ INAF -- Osservatorio Astronomico di Padova, Vicolo dell'Osservatorio 5, 
35122 Padova, Italy}

\pubyear{2004}
\volume{222}
\pagerange{1--8}
\date{?? and in revised form ??}
\jname{The Interplay among Black Holes, Stars and ISM \\in Galactic Nuclei}
\editors{Th. Storchi Bergmann, L.C. Ho \& H.R. Schmitt, eds.}

\begin{document}

\maketitle

\begin{abstract}
Near-infrared and optical imaging of BL Lac host galaxies is used to 
investigate their colour properties. We find that the $R$--$H$ colour and 
colour gradient distributions of the BL Lac hosts are much wider than those 
for normal ellipticals, and many objects have very blue hosts and/or steep 
colour gradients. The blue colours are most likely caused by recent 
star formation. The lack of obvious signs of interaction may, however, 
require a significant time delay between the interaction event with associated 
star formation episodes and the onset of the nuclear activity.
\end{abstract}

\firstsection 

\section{Introduction}

Optical imaging of low redshift (z $<$ 0.5) BL Lac objects 
(e.g. Falomo \& Kotilainen 1999; Urry et al. 2000) has shown that 
virtually all of them are hosted in luminous ellipticals, 
with characteristics indistinguishable from those of inactive 
massive ellipticals. On the other hand, near-infrared (NIR) imaging has only 
been available for small samples of BL Lacs (Kotilainen et al. 1998 [K98]; 
Scarpa et al. 2000 [S00]; Cheung et al. 2003 [C03]). We present deep 
high spatial resolution ($\sim$0.7 arcsec FWHM) NIR $H$--band (1.65 $\mu$m) 
imaging of 23 low redshift (z $<$ 0.3) BL Lacs that were previously 
investigated in the optical $R$-band (references above). We combine them with 
previous data to form a sample of 41 BL Lacs with which to investigate 
the optical-NIR colour properties of the BL Lac hosts and to compare 
them with radio galaxies (RG) and inactive ellipticals. Full report is given 
in Kotilainen \& Falomo (2004). 

\section{Host galaxy colours and colour gradients}

\begin{figure*}
  \begin{minipage}[t]{0.5\linewidth}
    \psfig{figure=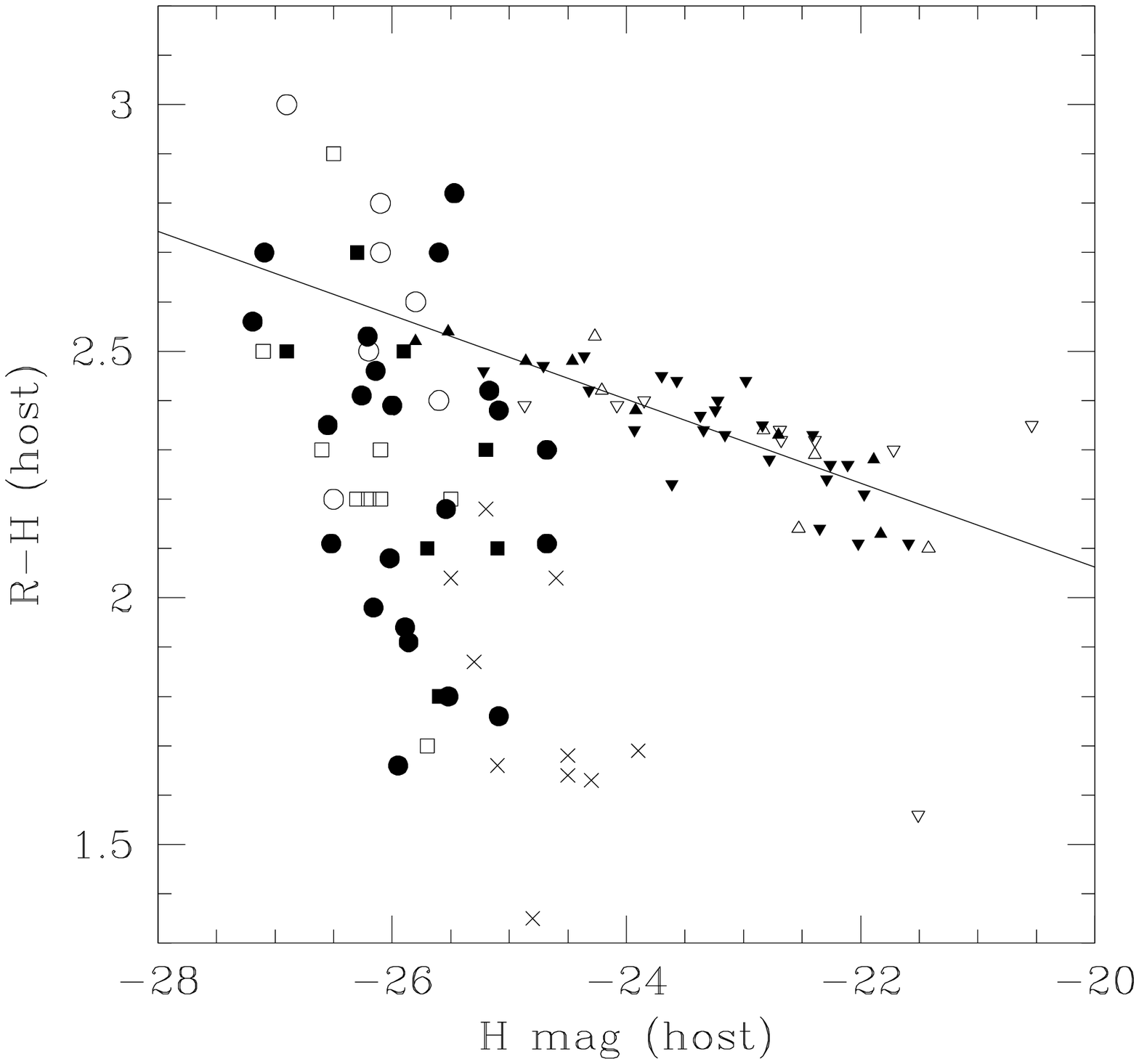,height=7cm}
  \end{minipage}
  \begin{minipage}[t]{0.5\linewidth}
    \psfig{figure=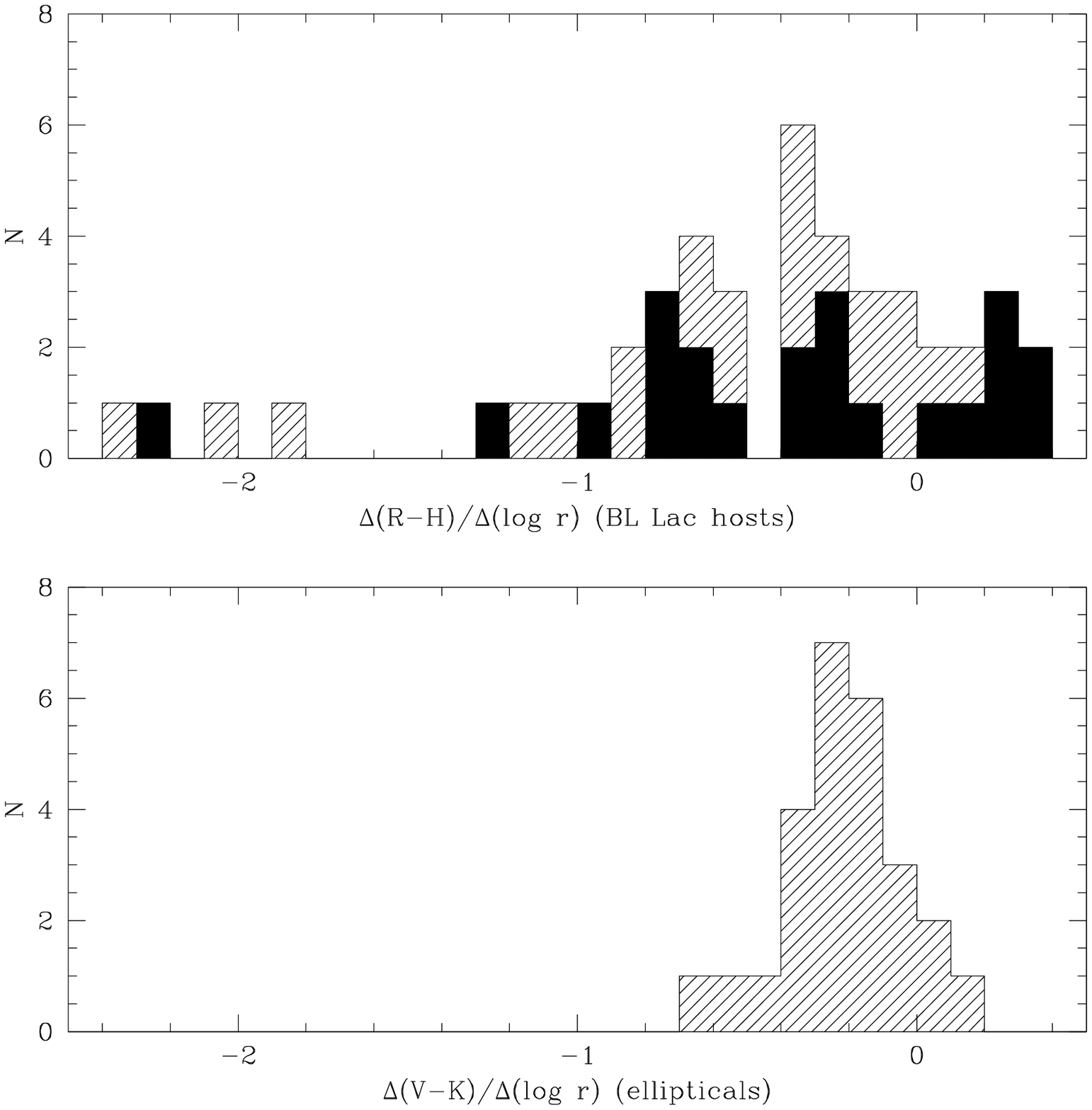,height=7cm}
  \end{minipage}
\caption{
{\bf Left:} The $R$--$H$ vs. $H$ C-M diagram for the BL Lac host galaxies in 
this work (filled circles), K98 (filled squares), S00 (open squares) and 
C03 (open circles). The other symbols indicate low redshift (z $<$ 0.2) 
elliptical quasar hosts (crosses; Jahnke et al. 2004) and elliptical galaxies 
in Virgo and Coma clusters (small triangles; Bower et al. 1992). 
The solid line shows the best--fit for the ellipticals.
{\bf Right:} Histogram of the $R$-$H$ colour gradient for the BL Lac hosts 
(top) and for inactive ellipticals (bottom; Peletier et al. 1990; 
Schombert et al. 1993). The solid histogram shows the BL Lacs in this work. 
}
\label{hrh}
\end{figure*}

It is well known that the colours of elliptical galaxies become bluer towards 
fainter magnitudes (e.g. Kodama \& Arimoto 1997). 
This colour--magnitude (C-M) relation may depend on age and metallicity.
Fig.~\ref{hrh} (left panel) shows 
the optical-NIR C-M relation of the BL Lac hosts, compared with ellipticals 
in clusters (Bower et al. 1992). BL Lac hosts exhibit a much broader range of 
colours, and they are systematically bluer than inactive ellipticals. 
This behaviour is similar to that found for 
low redshift RGs (Govoni et al. 2000) and 
quasar hosts (Jahnke et al. 2004). This trend could be explained if 
the C-M relation for ellipticals breaks down at the highest luminosities. 
Indeed, the BL Lac hosts cover the bright end of the luminosity function 
of ellipticals. However, although very few of the ellipticals are 
equally luminous, those few exhibit red colours. 
Therefore, if the C-M relation of ellipticals extends to high luminosity, 
the obvious explanation for the blue colours is recent star formation (SF). 
The broad range of colours then reflects differences in the SF epoch, 
from most recent SF episodes (blue) to old stellar populations (red). 

Inactive ellipticals have negative colour gradients (bluer with 
increasing radius) due to age and metallicity gradients 
(e.g. Tamura et al. 2000). The BL Lac hosts (and RGs; Govoni et al. 2000) 
show on average also a negative colour gradient (Fig.~\ref{hrh}, right panel), 
which is steeper and has a wider spread than that exhibited by 
inactive ellipticals (Peletier et al. 1990; Schombert et al. 1993). 
Intriguingly, we find a significant tail in the distribution with very strong 
negative gradients. These could be either intrinsic or due to central 
dust extinction in the host galaxies. Significant amount of dust in RGs has 
indeed been detected as dust lanes or dusty disks (Capetti et al. 2000).

If the blue colours are caused by a young stellar population component, 
they may signify the link between the interaction/merging event that triggered 
nuclear activity and rejuvenated the stellar population. However, optical and 
NIR images of BL Lac hosts show few obvious signs of interaction 
(e.g. tidal tails, asymmetries). This may necessitate a significant 
time delay ($>$ 100 Myr) between the SF episodes and the onset of 
the nuclear activity.

\end{document}